\documentclass[9pt,conference]{IEEEtran}

\usepackage[preprint]{dcase2025}

\usepackage{times,tabularx}
\usepackage{multirow}

\usepackage{acronym}
\usepackage{csquotes}
\usepackage{adjustbox}
\usepackage{balance}

\acrodef{asd}[ASD]{anomalous sound detection}
\acrodef{da}[DA]{domain adaptation}
\acrodef{dg}[DG]{domain generalization}
\acrodef{gmm}[GMM]{Gaussian mixture model}
\acrodef{smote}[SMOTE]{synthetic minority over-sampling technique}
\acrodef{vicreg}[VICreg]{variance-invariance-covariance regularization}

\title{Handling Domain Shifts for Anomalous Sound Detection:\\A Review of DCASE-Related Work}

\name{Kevin Wilkinghoff$\,^{1,2}$, Takuya Fujimura$^{3}$, Keisuke Imoto$^{4}$, Jonathan Le Roux$^{5}$, Zheng-Hua Tan$^{1,2}$, Tomoki Toda$^{3}$}

\address{$^{1}~$Aalborg University, Denmark, $^{2}~$Pioneer Centre for AI, Denmark, $^{3}~$Nagoya University, Japan\\
$^{4}~$Kyoto University, Japan, $^{5}~$Mitsubishi Electric Research Laboratories, MA, USA}

\begin{document}

\maketitle

\begin{abstract}
When detecting anomalous sounds in complex environments, one of the main difficulties is that trained models must be sensitive to subtle differences in monitored target signals, while many practical applications also require them to be insensitive to changes in acoustic domains. Examples of such domain shifts include changing the type of microphone or the location of acoustic sensors, which can have a much stronger impact on the acoustic signal than subtle anomalies themselves. Moreover, users typically aim to train a model only on source domain data, which they may have a relatively large collection of, and they hope that such a trained model will be able to generalize well to an unseen target domain by providing only a minimal number of samples to characterize the acoustic signals in that domain. In this work, we review and discuss recent publications focusing on this domain generalization problem for anomalous sound detection in the context of the DCASE challenges on acoustic machine condition monitoring.
\end{abstract}


\section{Introduction}
\Ac{asd} has a wide range of applications, including acoustic monitoring of machines \cite{kawaguchi2021description,dohi2022description,dohi2023description,nishida2024description,nishida2025description},
health \cite{dissanayake2021robust,murthy2021deep}, roads \cite{foggia2016audio},
smart home environments \cite{zieger2009acoustic},
or public places \cite{hayashi2018anomalous}.
The goal in all these applications is to distinguish between normal and anomalous audio recordings.
Usually, \ac{asd} systems are trained exclusively with normal data, as anomalous data is often difficult and costly to obtain.
\par
One of the major difficulties that \ac{asd} systems need to overcome is how to robustly handle so-called \textit{domain shifts}, changes in the recorded audio signal that are caused by changes in acoustic environments, sensors, or properties of monitored sound sources themselves.
Inherently, domain shifts have a strong impact on the audio signal and thus also affect the outcome of \ac{asd} systems if no precautions are taken.
However, in most applications, this effect is not desirable and should be suppressed.
Ideally, trained \ac{asd} systems should be completely insensitive to domain shifts, while being very sensitive to modifications of the monitored target signals that indicate the occurrence of application-dependent anomalies. 
As these differences may be very subtle compared to changes caused by domain shifts, especially in noisy environments with many sound sources, developing \ac{asd} systems that perform well in domain-shifted conditions is very challenging.
\par
\begin{figure}
    \centering
    \begin{adjustbox}{max width=\columnwidth}
          \includegraphics{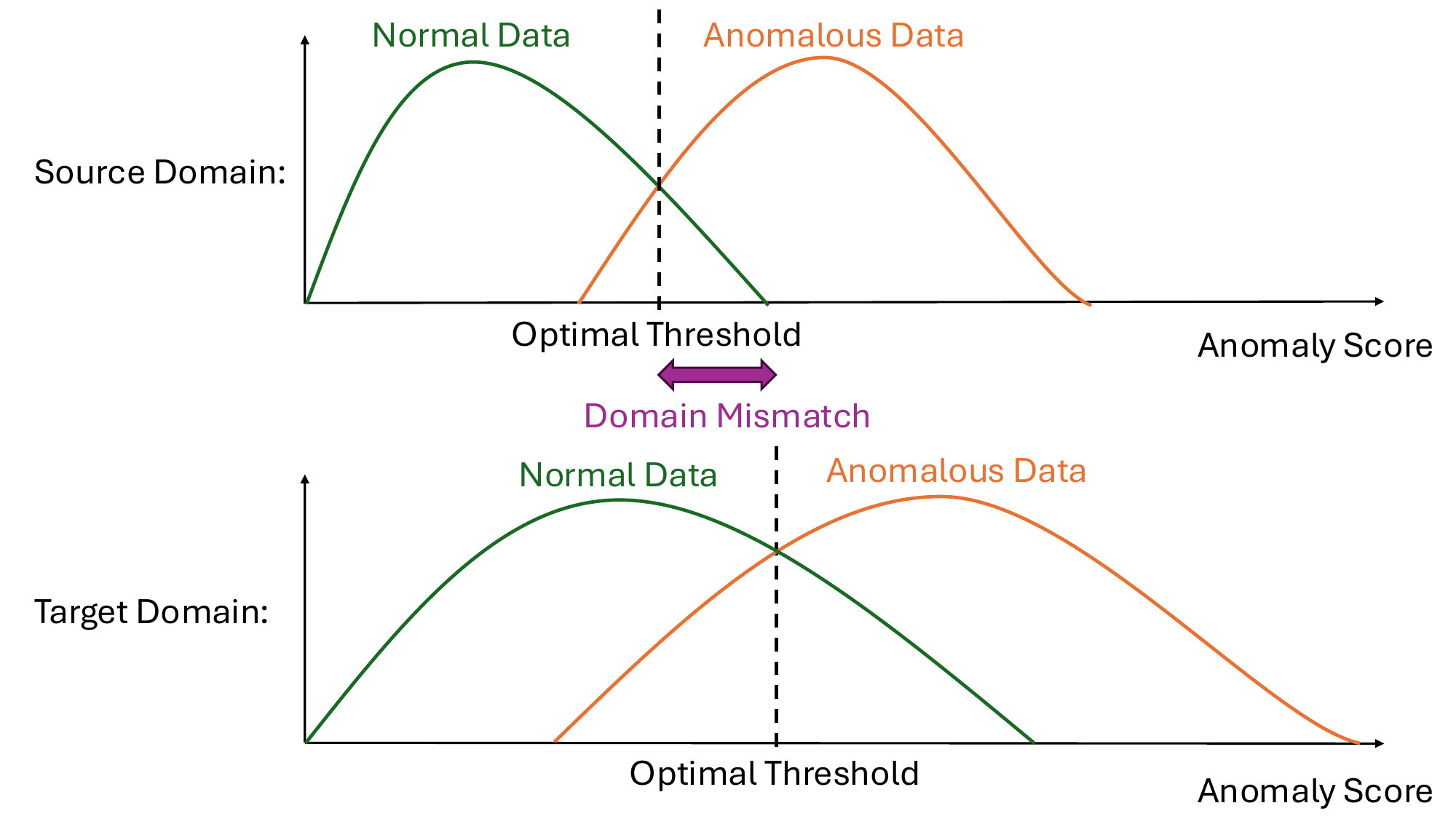}
    \end{adjustbox}
    \caption{Illustration of the domain mismatch between the anomaly scores of a source domain and a target domain.
    Normal and anomalous samples are usually less well separated in the target domain than in the source domain, which decreases domain-independent performance over the performance obtained for the source domain alone. Furthermore, the optimal decision thresholds for separating the scores belonging to the normal and anomalous data of different data domains differ substantially, which significantly decreases performance when using only a single threshold for both domains.
    }
    \label{fig:domain_mismatch}
\end{figure}

This paper, which extends a non-peer reviewed manuscript \cite{wilkinghoff2025handling}, reviews the latest works on how to handle domain shifts for \ac{asd} in the context of the annual DCASE challenge \cite{mesaros2024decade}.
We first describe the challenges posed by domain shifts and the resulting \textit{domain mismatch}, illustrated in \cref{fig:domain_mismatch}. We then describe the two main types of approaches to handling them, \textit{\ac{da}} and \textit{\ac{dg}}, defining the terms, listing publicly available datasets, and reviewing recently published works related to these topics.
Last but not least, the limitations of different approaches are discussed and future directions of research are identified.

\begin{table*}[t]
	\centering
    \sisetup{
    detect-weight, 
    mode=text, 
    tight-spacing=true,
    round-mode=places,
    round-precision=0,
    table-format=3,
    table-number-alignment=center
}
	\caption{DCASE \ac{asd} datasets in domain-shifted conditions.}
\begin{adjustbox}{max width=0.905\textwidth}
	\begin{tabular}{llS[table-format=2]ccccccSSSSS}
		\toprule
        &&&&&&&&&\multicolumn{5}{c}{\# recordings (per section)}\\
        \cmidrule(lr){10-14}
        &&\multicolumn{3}{c}{\# machine types}&\multicolumn{3}{c}{\# sections (per machine type)}&&\multicolumn{2}{c}{source domain}&\multicolumn{2}{c}{target domain}&\\
        \cmidrule(lr){3-5}\cmidrule(lr){6-8}\cmidrule(lr){10-11}\cmidrule(lr){12-13}
        Name&Purpose&{total}&dev.\ set&eval.\ set&total&dev.\ set&eval.\ set&split&{normal}&{anomalous}&{normal}&{anomalous}&{supplemental}\\
		\midrule
		\multirow{2}{*}{DCASE2021 \cite{kawaguchi2021description}} & \multirow{2}{*}{\ac{da}} & {\multirow{2}{*}{7}} & \multirow{2}{*}{7} & \multirow{2}{*}{7} & \multirow{2}{*}{6} & \multirow{2}{*}{3} & \multirow{2}{*}{3} & train & 1000 & 0 & 3 & 0 & 0\\
		&&&&&&&& test & 100 & 100 & 100 & 100 & 0\\
		\midrule
		\multirow{2}{*}{DCASE2022 \cite{dohi2022description}} & \multirow{2}{*}{\ac{dg}} & {\multirow{2}{*}{7}} & \multirow{2}{*}{7} & \multirow{2}{*}{7} & \multirow{2}{*}{6} & \multirow{2}{*}{3} & \multirow{2}{*}{3} & train & 990 & 0 & 10 & 0 & 0\\
		&&&&&&&& test & 50 & 50 & 50 & 50 & 0\\
        \midrule
		\multirow{2}{*}{DCASE2023 \cite{dohi2023description}} & \multirow{2}{*}{\ac{dg}} & {\multirow{2}{*}{14}} & \multirow{2}{*}{7} & \multirow{2}{*}{7} & \multirow{2}{*}{1} & \multirow{2}{*}{1} & \multirow{2}{*}{1} & train & 990 & 0 & 10 & 0 & 0\\
		&&&&&&&& test & 50 & 50 & 50 & 50 & 0\\
        \midrule
		\multirow{2}{*}{DCASE2024 \cite{nishida2024description}} & \multirow{2}{*}{\ac{dg}} & {\multirow{2}{*}{16}} & \multirow{2}{*}{7} & \multirow{2}{*}{9} & \multirow{2}{*}{1} & \multirow{2}{*}{1} & \multirow{2}{*}{1} & train & 990 & 0 & 10 & 0 & 0\\
		&&&&&&&& test & 50 & 50 & 50 & 50 & 0\\
        \midrule
		\multirow{2}{*}{DCASE2025 \cite{nishida2025description}} & \multirow{2}{*}{\ac{dg}} & {\multirow{2}{*}{14}} & \multirow{2}{*}{7} & \multirow{2}{*}{7} & \multirow{2}{*}{1} & \multirow{2}{*}{1} & \multirow{2}{*}{1} & train & 990 & 0 & 10 & 0 & 100\\
		&&&&&&&& test & 50 & 50 & 50 & 50 & 0\\
		\bottomrule
	\end{tabular}
\end{adjustbox}
\label{tab:datasets}
\end{table*}

\section{Domain Shifts}

The goal of \ac{asd} is to determine, based on a sound recording of a phenomenon, whether that phenomenon is normal or anomalous.
In the context of the DCASE Challenges, the objective is to differentiate between normal and anomalous sounds produced by a known type of machine.
The normal/anomalous characteristic of a phenomenon is an intrinsic property of that phenomenon: as such, the determination by the \ac{asd} system should ideally be independent of the conditions in which the sound was recorded, as long as the change in conditions does not alter the observability and nature of the phenomenon's normal/anomalous characteristic. Such changes include the use of microphones of different types or in different locations, the presence or absence of other sound sources, or modifications in certain properties of the monitored sound sources themselves, e.g., the use of different settings of the monitored machines.
\par
\ac{asd} systems are however often placed in a conundrum: their only view of the phenomenon's normal/anomalous characteristic is via a training set of normal data recorded under a certain set of conditions, referred to as a \textit{source domain}, and they are expected to make a determination on the normal/anomalous characteristic on a new sample recorded under a potentially different set of conditions, referred to as a \textit{target domain}, for which they only have a few training samples. Changes in the recording conditions occurring between the source domain and the target domain constitute a so-called \textit{domain shift}.
\ac{asd} systems thus need to implicitly disentangle the effects on the sound signal of the normal/anomalous characteristic and the recording conditions, in order to be robust to changes in these conditions.
\par
An additional difficulty lies in the large imbalance often observed in practice between the number of available training samples in the source and target domains. Many training samples are typically available for the source domain, while only a few training samples may be available for the target domain. 
Recording a large amount of new training samples after a domain shift occurred may be impractical, as it requires much effort, and thus is very costly; domain shifts may also frequently or even continuously occur; and operators may not even be aware of such domain shifts.
\par
A consequence of domain shifts related to differences in signal space is that anomaly scores are usually also distributed very differently across both domains.
This results in a so-called \textit{domain mismatch}, illustrated in \cref{fig:domain_mismatch}.
Since the embedding models were trained with no or few samples belonging to the target domain, the anomaly score distributions for the normal and anomalous samples are not well-separated in the target domain, which inherently leads to a worse \ac{asd} performance than in the source domain.
Furthermore, the optimal decision thresholds differ substantially, further degrading the performance if the same decision threshold is used for both domains.

\section{Domain Adaptation}
One way to handle domain shifts is to adapt an existing \ac{asd} system that was trained with sufficient data from a source domain to a specific target domain.
Here, the main challenge is that the training set for the target domain consists of only very few samples and thus knowledge from the source domain, which may differ substantially from the target domain, needs to be transferred somehow to obtain a well-performing system.
Note that once a system is adapted to a target domain, it may no longer perform well in the source domain it was trained on.

\subsection{Datasets}
A dataset focusing on handling domain shifts for \ac{asd} through \ac{da} is the DCASE2021 \ac{asd} dataset \cite{kawaguchi2021description}.
This dataset contains \SI{10}{\second} recordings of five different machine types from MIMII DUE \cite{tanaba2021mimiii_due} and two additional machine types from ToyADMOS2 \cite{harada2021toyadmos2} that are combined with background noise from real factories.
The dataset is divided into a development set and an evaluation set, which both contain three sections for each machine type.
These sections are specific partitions of the dataset for calculating the performance and may contain recordings from multiple machines of the same type.
Both the development and evaluation sets consist of a training split and a test split.
The training splits contain only normal data, of which $1{,}000$ samples belong to the source domain and only $3$ samples belong to the target domain, resulting in a very imbalanced dataset in terms of domain.
For each training file, additional attribute information about machine settings or the acoustic environment is provided, and it can be utilized for training \ac{asd} systems.
The test splits contain $200$ samples for each domain, of which one half are normal and the other half are anomalous.
For each test file, it is known whether these files belong to the source or the target domain, but it is unknown whether they are normal or anomalous.
Summarized statistics of this dataset can be found in the top part of \cref{tab:datasets}.

\subsection{Approaches}
\begin{table*}
	\centering
	\caption{\Ac{asd} approaches for handling domain shifts.}
\begin{adjustbox}{max width=\textwidth}
	\begin{tabular}{llrlc}
		\toprule
        Topic&Strategy&\multicolumn{1}{c}{Reference}&Main idea&DCASE datasets used\\
		\midrule
        \ac{da}&-&Chen et al \cite{chen2022self}&fine-tune model&2021\\
        \ac{da}&-&Kuroyanagi et al \cite{kuroyanagi2021ensemble}&fine-tune models and domain-specific backends&2021\\
        \ac{da}&-&Yamaguchi et al \cite{yamaguchi2019adaflow}&fine-tune batch normalization layers&other\\
        \ac{da}&-&Lopez et al \cite{lopez2021ensemble}&domain-specific normalization with AutoDIAL&2021\\
        \ac{da}&-&Chen et al \cite{chen2022learning}&gradient-based meta learning&2021\\
        \ac{da}&-&Wilkinghoff \cite{wilkinghoff2021combining}&domain-specific backends&2021\\
		\midrule
        \ac{dg}&domain specialization&Kuroyanagi et al \cite{kuroyanagi2022two}&domain-specific models and domain classifier&2022\\
        \ac{dg}&domain specialization&Kuroyanagi et al \cite{kuroyanagi2022two}&balance mini-batches with mixup&2022\\
        \ac{dg}&domain specialization&Guan et al \cite{guan2023time}&balance domains with \acs{smote}&2022\\
        \ac{dg}&domain specialization&Junjie et al \cite{junjie2023anomaly}&balance domains with \acs{smote}&2023\\
        \ac{dg}&domain specialization&Chen et al \cite{chen2025multi-scale}&balance domains with \acs{smote} and a mixup variant&2022,2023\\
        \midrule
        \ac{dg}&domain-invariant representations&Deng et al \cite{deng2022ensemble}&normalize samples independently for each batch&2022\\
        \ac{dg}&domain-invariant representations&Nejjar et al \cite{nejjar2022dg-mix}&\acs{vicreg} extension with mixup&2022\\
        \ac{dg}&domain-invariant representations&Yan et al \cite{yan2024transformer}&minimize covariance differences between domains&2020, 2022\\
        \midrule
        \ac{dg}&feature disentanglement&Venkatesh et al \cite{venkatesh2022improved}&discriminative tasks for sections and attributes&2022\\
        \ac{dg}&feature disentanglement&Dohi et al \cite{dohi2022disentangling}&disentangle attributes with normalizing flow&other\\
        \ac{dg}&feature disentanglement&Lan et al \cite{lan2024hierarchical}& Mahalanobis distances for hierarchical metadata&2022\\
        \ac{dg}&feature disentanglement&Guan et al \cite{guan2025disentangling}&gradient reversal with hierarchical metadata&2022\\
        \midrule
        \ac{dg}&anomaly score calculation&Harada et al \cite{harada2023first-shot}&domain-specific Mahalanobis distance&2022\\
        \ac{dg}&anomaly score calculation&Wilkinghoff \cite{wilkinghoff2023design}&nearest neighbor based anomaly scores&2022\\
        \ac{dg}&anomaly score calculation&Wilkinghoff et al \cite{wilkinghoff2025local}&local density based score normalization&2020, 2023, 2024\\
        \ac{dg}&anomaly score calculation&Saengthong et al \cite{saengthong2024deep}&domain-wise standardization of scores&2020, 2023\\
		\bottomrule
	\end{tabular}
\end{adjustbox}
\label{tab:approaches}
\end{table*}
Approaches for \ac{da} mostly focus on first training an \ac{asd} system with data from the source domain and then adapting this trained system to a specific target domain.
One possibility to do this is to fine-tune an entire model that was trained in the source domain with data belonging to the target domain \cite{kuroyanagi2021ensemble,chen2022self}.
This changes the problem of balancing the very differently sized training datasets of both domains to a problem of preventing the model from overfitting to the few target domain samples available for training.
In \cite{yamaguchi2019adaflow}, only the parameters of batch normalization layers \cite{ioffe2015batch} are fine-tuned to the target domain to minimize the computational costs needed for the adaptation while also reducing overfitting effects.
A similar idea based on domain-specific normalization layers is realized in \cite{lopez2021ensemble} by using AutoDIAL \cite{carluzzi2017autodial}.
The authors of \cite{chen2022learning} propose to use gradient-based meta learning \cite{nichol2018first} and a prototypical loss \cite{snell2017prototypical} in order to more effectively adapt to target domains with only a very small amount of training samples.
Another \ac{da} approach is to simply train a joint embedding model for both domains but estimate the distributions of each domain individually \cite{wilkinghoff2021combining,kuroyanagi2021ensemble}.
\Cref{tab:approaches} briefly summarizes all mentioned approaches.

\section{Domain Generalization}
Adapting models for each new domain with the possible need of re-training models, fine-tuning hyperparameters, or even replacing system components is very impractical, as it is computationally costly and may require expert knowledge.
It is thus much more desirable to strive for {\acf{dg}} \cite{wang2023generalizing,zhou2023domain}, with the goal of obtaining a model that performs well in the source domain and generalizes well to unseen target domains by only providing a few samples to define how normal sounds belonging to these domains should sound like.
Since \ac{dg} is literally a generalization of \ac{da} to arbitrary domains rather than a specific target domain, it is more difficult to obtain such a model.
To highlight the practical differences between \ac{da} and \ac{dg}, the reader is reminded that for \ac{da}, models can be fully adapted to a specific target domain, even if performance in the source domain is weak after adaptation.
In contrast, \ac{dg} requires strong performance in the source and target domain, using even the same decision threshold for the anomaly scores in both domains.

\subsection{Datasets}
Currently, there are four DCASE \ac{asd} datasets focusing on \ac{dg}:
the DCASE2022 \cite{dohi2022description}, DCASE2023 \cite{dohi2023description}, DCASE2024 \cite{nishida2024description}, and DCASE2025 \cite{nishida2025description} datasets, which are all based on MIMII DG \cite{dohi2022mimiidg} as well as, respectively, ToyADMOS2 \cite{harada2021toyadmos2}, ToyADMOS2+ \cite{harada2023toyadmos2+}, and ToyADMOS2\# \cite{niizumi2024toyadmos2sharp}.
A part of the DCASE2024 dataset was recorded with the same setup as IMAD-DS \cite{albertini2024imad-ds}.
IMAD-DS is another \ac{asd} dataset containing noisy multi-sensor signals of two machines in domain-shifted conditions recorded with a microphone, an accelerometer, and a gyroscope.
The fundamental difference with the DCASE2021 dataset, which focuses on \ac{da}, is that the domain of individual test samples is unknown during inference.
Moreover, the performance for each section is computed jointly for the source and target domains, i.e., a single decision threshold needs to be used for both domains.
Other less severe differences include the size of all test sets, which is only half the size of the DCASE2021 test sets, and the availability of $10$ normal training samples belonging to the target domain instead of $3$.
The differences between the DCASE2021 and DCASE2022 datasets end there, making them relatively similar.
In contrast, the DCASE2023, DCASE2024, and DCASE2025 datasets only consist of a single section for each machine type, and the development and evaluation sets contain recordings of completely different machine types.
Furthermore, the DCASE2024 and DCASE2025 datasets have noise conditions that are exclusively used for specific machine types, and for some machine types no additional attribute information is provided.
The major difference between the DCASE2024 and DCASE2025 datasets is that supplemental recordings containing noise-free recordings of normal machine sounds or only background noise are provided for the train splits of the DCASE2025 dataset.
Each year, modifications to the datasets result in a more realistic yet more challenging \ac{asd} task.
For more details about these datasets, the reader is referred to \cref{tab:datasets} and the corresponding references.

\subsection{Approaches}
\ac{dg} approaches for \ac{asd} in the context of DCASE can be grouped into four different strategies, which will now be discussed.
\cref{tab:approaches} contains an overview of these approaches.

\subsubsection{Domain Specialization}
A simple approach for reducing domain mismatch is to balance the number of training samples belonging to the source and target domains.
This can be achieved by balancing the domains in each mini-batch \cite{kuroyanagi2022two} or using more sophisticated approaches \cite{guan2023time,junjie2023anomaly,chen2025multi-scale} such as \ac{smote} \cite{chawla2002smote}.
However, since this approach trains models for specific domain shifts and thus requires to re-train the entire \ac{asd} system for each possible domain shift, this can be referred to as \textit{weak \ac{dg}} or \textit{domain specialization}.
Note that in contrast to \ac{da}, \ac{asd} systems need to work well for all domains without knowledge about the domain a given sample belongs to.
Another way to handle this is to train domain-specific models and a domain classifier \cite{kuroyanagi2022two}.

\subsubsection{Domain-invariant Representations}
The idea of domain-invariant representation learning \cite{ben-david2006analysis} or domain-mixing-based approaches \cite{dohi2022description} is to reduce the variance between multiple domains based on the assumption that this also reduces the variance to other arbitrary target domains.
In \cite{deng2022ensemble}, individual samples of a batch are normalized independently of each other instead of using batch normalization \cite{ioffe2015batch} to avoid learning statistics that over-emphasize the source domain due to the highly imbalanced number of samples.
The authors of \cite{nejjar2022dg-mix} propose DG-mix, an extension of \ac{vicreg} \cite{bardes2022vicreg} for self-supervised pre-training, using a loss term that minimizes the difference between domains and virtual domains created by mixup \cite{zhang2017mixup} before fine-tuning the model.
Motivated by \cite{sun2016deep}, which reduces the difference between second-order statistics of intermediate representations from domain-specific convolutional neural networks, \cite{yan2024transformer} aims at reducing the difference between second-order statistics of source domain features and target domain features.
To this end, the authors applied augmentations such as pitch shifting, time shifting, time stretching, adding white noise, and Filteraugment \cite{nam2022filteraugment} to target domain samples to create more diverse domain shifts for training.

\subsubsection{Feature Disentanglement}
The main idea of feature disentanglement \cite{zhang2022towards} is to decompose representations into domain-related features that are invariant for different domains and domain-unrelated features that capture the variability of different domains.
The latter are independent of the domain and thus also generalize well to unseen domains.
A possible way to achieve this is to use a domain-classification-based approach \cite{dohi2022description}.
In most works on \ac{asd} in the DCASE context, this is achieved by focusing on the provided attribute information.
The authors of \cite{venkatesh2022improved} use two discriminative tasks for the sections and attributes. In \cite{dohi2022disentangling}, attribute information is disentangled in a normalizing-flow-based \ac{asd} model.
In \cite{lan2024hierarchical}, a combination of a hierarchical metadata structure and attribute-specific Mahalanobis distances is used to learn more domain-related features.
This approach is extended by \cite{guan2025disentangling} with gradient-reversal-based feature disentanglement of attribute information \cite{ganin2016domain} and the use of a focal loss \cite{lin2017focal}.

\subsubsection{Anomaly Score Calculation}
Last but not least, \ac{dg} capabilities can also be improved by modifying the anomaly score computation.
This approach has the advantage of not requiring expensive re-training of the neural networks that serve as the basis of state-of-the-art \ac{asd} systems. 
To this end, \cite{harada2023first-shot} uses an autoencoder and calculates the Mahalanobis distance between input data and reconstruction using domain-specific covariance matrices.
For discriminative \ac{asd} systems, it was shown that simple nearest-neighbor-based anomaly scores lead to better results than estimating domain-specific distributions \cite{wilkinghoff2023design}.
Further improvements can be obtained by normalizing the anomaly scores to reduce the domain mismatch.
Examples are normalizing the anomaly scores based on local densities of normal reference samples \cite{wilkinghoff2025keeping,wilkinghoff2025local} or by a domain-wise standardization of the anomaly score distributions \cite{saengthong2024deep}.

\section{Limitations and Future Directions}
\begin{table}
	\centering
	\caption{Comparison of different types of \acs{dg} approaches.}
\begin{adjustbox}{max width=\columnwidth}
	\begin{tabular}{lccc}
		\toprule
        \acs{dg} approach & Effort & Requires labels & Effectiveness\\
		\midrule
        Domain Specialization & \cellcolor{red!20} high & \cellcolor{red!20} yes & \cellcolor{green!30!white} high\\
        Domain-invariant Representations & \cellcolor{yellow!30!white} medium & \cellcolor{green!30!white} no & \cellcolor{yellow!30!white} medium\\
        Feature Disentanglement & \cellcolor{green!15!white} low & \cellcolor{red!20} yes & \cellcolor{yellow!30!white} medium\\
        Anomaly Score Calculation & \cellcolor{green!30!white} very low & \cellcolor{yellow!30!white} sometimes & \cellcolor{red!20} low\\
		\bottomrule
	\end{tabular}
\end{adjustbox}
\label{tab:dg_comparison}
\end{table}
All presented \ac{dg} approaches have limitations.
A comparison of the approaches can be found in \cref{tab:dg_comparison}.
Although domain specialization approaches are highly effective, they can only adapt to specific domains and thus the entire \ac{asd} system, including the embedding model, needs to be re-trained for each domain shift, which is highly impractical as it is a strenuous process requiring access to domain labels.
Obtaining domain-invariant representations does not require domain labels, but requires to simulate a diverse set of realistic domain shifts, which is a difficult task by itself and thus may limit the effectiveness.
Feature disentanglement does not require much additional effort, but does require access to labeled data of various domains to be effective, which may not always be available.
Focusing only on the calculation of the anomaly scores usually does not require one to modify the training of the embedding models.
However, this approach is still sensitive to domain shifts if these also affect the obtained embeddings, thus limiting its effectiveness, even when making use of domain labels.
Depending on the application and the available data, different \ac{dg} approaches can be jointly used to mitigate the limitations of individual approaches.
For example, anomaly score calculation approaches for \ac{dg} can be used regardless of the learned representations, and simulated domain shifts, as used when learning domain-invariant representations, can also be used for feature disentanglement.
However, the underlying issues of individual approaches persist and future work on \ac{dg} is needed to solve these problems and identify potential synergies between specific approaches.
\par
A future direction of research is continuous \ac{da} \cite{wang2020continuously}.
Here, the discrete indices for the domains are replaced with continuous ones to model the underlying relations between different domains.
This improves consistency and may even allow to handle domains for which no training samples are available.
Another direction is continual \ac{da} \cite{wang2022continual} that aims at handling domain shifts in non-stationary environments where domains are not static and may change continuously over time.
In these environments, only adapting to a fixed domain is not sufficient.
Note that continuous domain indices and continuously adapting to non-static domains can also be realized for \ac{dg}.
As a third direction, explaining the decisions of \ac{asd} systems is an open problem \cite{mai2022explaining,tsubaki2023audio}, particularly in domain-shifted conditions.
Here, future work may focus on explaining decisions of \ac{asd} in unknown target domains \cite{gao2024learning} or explaining domain shifts themselves \cite{bobek2023towards}.
Explainable \ac{asd} systems that focus on particular characteristics of the data may even offer better \ac{dg} capabilities \cite{zunino2021explainable}.
Last but not least, the semi-supervised \ac{asd} setting in the DCASE context can be relaxed to an unsupervised setting where even the training dataset may contain unlabeled anomalies \cite{aggarwal2017outlier}.

\section{Conclusion}
In this work, we reviewed recent work on how to handle domain shifts for \ac{asd} tasks related to DCASE.
We first defined domain shifts in the context of \ac{asd}, then discussed the topics of \textit{\acf{da}} and \textit{\acf{dg}} by motivating and defining these terms, presenting relevant datasets, and collecting different works related to these topics.
In particular, \ac{dg} approaches were grouped into \textit{domain specialization}, \textit{domain-invariant representations}, \textit{feature disentanglement}, and \textit{anomaly score calculation}.
Furthermore, \textit{continuous} and \textit{continual \ac{da}} as well as \textit{explainable \ac{dg}} were identified as possible future research directions.
For future work, we plan to compare all presented techniques with ASDKit \cite{fujimura2025asdkit} through extensive experiments to identify the best approaches and potential synergies.


\bibliographystyle{IEEEtran}
\bibliography{refs}

\begin{thebibliography}{10}
\providecommand{\url}[1]{#1}
\csname url@samestyle\endcsname
\providecommand{\newblock}{\relax}
\providecommand{\bibinfo}[2]{#2}
\providecommand{\BIBentrySTDinterwordspacing}{\spaceskip=0pt\relax}
\providecommand{\BIBentryALTinterwordstretchfactor}{4}
\providecommand{\BIBentryALTinterwordspacing}{\spaceskip=\fontdimen2\font plus
\BIBentryALTinterwordstretchfactor\fontdimen3\font minus \fontdimen4\font\relax}
\providecommand{\BIBforeignlanguage}[2]{{%
\expandafter\ifx\csname l@#1\endcsname\relax
\typeout{** WARNING: IEEEtran.bst: No hyphenation pattern has been}%
\typeout{** loaded for the language `#1'. Using the pattern for}%
\typeout{** the default language instead.}%
\else
\language=\csname l@#1\endcsname
\fi
#2}}
\providecommand{\BIBdecl}{\relax}
\BIBdecl

\bibitem{kawaguchi2021description}
Y.~Kawaguchi, K.~Imoto, Y.~Koizumi, N.~Harada, D.~Niizumi, K.~Dohi, R.~Tanabe, H.~Purohit, and T.~Endo, ``Description and discussion on {DCASE} 2021 challenge task 2: {U}nsupervised anomalous detection for machine condition monitoring under domain shifted conditions,'' in \emph{Proc. DCASE}, 2021.

\bibitem{dohi2022description}
K.~Dohi, K.~Imoto, N.~Harada, D.~Niizumi, Y.~Koizumi, T.~Nishida, H.~Purohit, T.~Endo, M.~Yamamoto, and Y.~Kawaguchi, ``Description and discussion on {DCASE} 2022 challenge task 2: {U}nsupervised anomalous sound detection for machine condition monitoring applying domain generalization techniques,'' in \emph{Proc. DCASE}, 2022.

\bibitem{dohi2023description}
K.~Dohi, K.~Imoto, N.~Harada, D.~Niizumi, Y.~Koizumi, T.~Nishida, H.~Purohit, R.~Tanabe, T.~Endo, and Y.~Kawaguchi, ``Description and discussion on {DCASE} 2023 challenge task 2: {F}irst-shot unsupervised anomalous sound detection for machine condition monitoring,'' in \emph{Proc. DCASE}, 2023.

\bibitem{nishida2024description}
T.~Nishida, N.~Harada, D.~Niizumi, D.~Albertini, R.~Sannino, S.~Pradolini, F.~Augusti, K.~Imoto, K.~Dohi, H.~Purohit, T.~Endo, and Y.~Kawaguchi, ``Description and discussion on {DCASE} 2024 challenge task 2: {F}irst-shot unsupervised anomalous sound detection for machine condition monitoring,'' in \emph{Proc. DCASE}, 2024.

\bibitem{nishida2025description}
------, ``Description and discussion on {DCASE} 2025 challenge task 2: First-shot unsupervised anomalous sound detection for machine condition monitoring,'' \emph{arXiv preprint arXiv:2506.10097}, 2025.

\bibitem{dissanayake2021robust}
T.~Dissanayake, T.~Fernando, S.~Denman, S.~Sridharan, H.~Ghaemmaghami, and C.~Fookes, ``A robust interpretable deep learning classifier for heart anomaly detection without segmentation,'' \emph{{IEEE} J. Biomed. Health Informatics}, vol.~25, no.~6, 2021.

\bibitem{murthy2021deep}
S.~N. Murthy and E.~Agu, ``Deep learning anomaly detection methods to passively detect {COVID-19} from audio,'' in \emph{Proc. ICDH}, 2021.

\bibitem{foggia2016audio}
P.~Foggia, N.~Petkov, A.~Saggese, N.~Strisciuglio, and M.~Vento, ``Audio surveillance of roads: {A} system for detecting anomalous sounds,'' \emph{{IEEE} Trans. Intell. Transp. Syst.}, vol.~17, no.~1, 2016.

\bibitem{zieger2009acoustic}
C.~Zieger, A.~Brutti, and P.~Svaizer, ``Acoustic based surveillance system for intrusion detection,'' in \emph{Proc. AVSS}, 2009.

\bibitem{hayashi2018anomalous}
T.~Hayashi, T.~Komatsu, R.~Kondo, T.~Toda, and K.~Takeda, ``Anomalous sound event detection based on {W}ave{N}et,'' in \emph{Proc. EUSIPCO}, 2018.

\bibitem{wilkinghoff2025handling}
K.~Wilkinghoff, T.~Fujimura, K.~Imoto, and J.~Le~Roux, ``Handling domain shifts for anomalous sound detection: A review,'' in \emph{Proc. DAS/DAGA}, 2025.

\bibitem{mesaros2024decade}
A.~Mesaros, R.~Serizel, T.~Heittola, T.~Virtanen, and M.~D. Plumbley, ``A decade of {DCASE:} achievements, practices, evaluations and future challenges,'' in \emph{Proc. ICASSP}, 2025.

\bibitem{tanaba2021mimiii_due}
R.~Tanabe, H.~Purohit, K.~Dohi, T.~Endo, Y.~Nikaido, T.~Nakamura, and Y.~Kawaguchi, ``{MIMII} {DUE}: {S}ound dataset for malfunctioning industrial machine investigation and inspection with domain shifts due to changes in operational and environmental conditions,'' in \emph{Proc. WASPAA}, 2021.

\bibitem{harada2021toyadmos2}
N.~Harada, D.~Niizumi, D.~Takeuchi, Y.~Ohishi, M.~Yasuda, and S.~Saito, ``Toy{ADMOS2:} {A}nother dataset of miniature-machine operating sounds for anomalous sound detection under domain shift conditions,'' in \emph{Proc. DCASE}, 2021.

\bibitem{chen2022self}
H.~Chen, Y.~Song, L.~Dai, I.~McLoughlin, and L.~Liu, ``Self-supervised representation learning for unsupervised anomalous sound detection under domain shift,'' in \emph{Proc. ICASSP}, 2022.

\bibitem{kuroyanagi2021ensemble}
I.~Kuroyanagi, T.~Hayashi, Y.~Adachi, T.~Yoshimura, K.~Takeda, and T.~Toda, ``An ensemble approach to anomalous sound detection based on conformer-based autoencoder and binary classifier incorporated with metric learning,'' in \emph{Proc. DCASE}, 2021.

\bibitem{yamaguchi2019adaflow}
M.~Yamaguchi, Y.~Koizumi, and N.~Harada, ``{AdaFlow}: {D}omain-adaptive density estimator with application to anomaly detection and unpaired cross-domain translation,'' in \emph{Proc. ICASSP}, 2019.

\bibitem{lopez2021ensemble}
J.~A. Lopez, G.~Stemmer, P.~Lopez{-}Meyer, P.~Singh, J.~A. del Hoyo~Ontiveros, and H.~A. Cordourier, ``Ensemble of complementary anomaly detectors under domain shifted conditions,'' in \emph{Proc. DCASE}, 2021.

\bibitem{chen2022learning}
B.~Chen, L.~Bondi, and S.~Das, ``Learning to adapt to domain shifts with few-shot samples in anomalous sound detection,'' in \emph{Proc. ICPR}, 2022.

\bibitem{wilkinghoff2021combining}
K.~Wilkinghoff, ``Combining multiple distributions based on sub-cluster {AdaCos} for anomalous sound detection under domain shifted conditions,'' in \emph{Proc. DCASE}, 2021.

\bibitem{kuroyanagi2022two}
I.~Kuroyanagi, T.~Hayashi, K.~Takeda, and T.~Toda, ``Two-stage anomalous sound detection systems using domain generalization and specialization techniques,'' in \emph{Proc. DCASE}, 2022.

\bibitem{guan2023time}
J.~Guan, Y.~Liu, Q.~Zhu, T.~Zheng, J.~Han, and W.~Wang, ``Time-weighted frequency domain audio representation with {GMM} estimator for anomalous sound detection,'' in \emph{Proc. ICASSP}, 2023.

\bibitem{junjie2023anomaly}
W.~Junjie, W.~Jiajun, C.~Shengbing, S.~Yong, and L.~Mengyuan, ``Anomaly sound detection system based on multi-dimensional attention module,'' {DCASE2023} Challenge, Tech. Rep., 2023.

\bibitem{chen2025multi-scale}
S.~Chen, Y.~Sun, J.~Wang, M.~Wan, M.~Liu, and X.~Li, ``A multi-scale dual-decoder autoencoder model for domain-shift machine sound anomaly detection,'' \emph{Digit. Signal Process.}, vol. 156, 2025.

\bibitem{deng2022ensemble}
Y.~Deng, A.~Jiang, Y.~Duan, J.~Ma, X.~Chen, J.~Liu, P.~Fan, C.~Lu, and W.~Zhang, ``Ensemble of multiple anomalous sound detectors,'' in \emph{Proc. DCASE}, 2022.

\bibitem{nejjar2022dg-mix}
I.~Nejjar, J.~Meunier{-}Pion, G.~Frusque, and O.~Fink, ``{DG-Mix: D}omain generalization for anomalous sound detection based on self-supervised learning,'' in \emph{Proc. DCASE}, 2022.

\bibitem{yan2024transformer}
J.~Yan, Y.~Cheng, Q.~Wang, L.~Liu, W.~Zhang, and B.~Jin, ``Transformer and graph convolution-based unsupervised detection of machine anomalous sound under domain shifts,'' \emph{{IEEE} Trans. Emerg. Top. Comput. Intell.}, vol.~8, no.~4, 2024.

\bibitem{venkatesh2022improved}
S.~Venkatesh, G.~Wichern, A.~S. Subramanian, and J.~{Le Roux}, ``Improved domain generalization via disentangled multi-task learning in unsupervised anomalous sound detection,'' in \emph{Proc. DCASE}, 2022.

\bibitem{dohi2022disentangling}
K.~Dohi, T.~Endo, and Y.~Kawaguchi, ``Disentangling physical parameters for anomalous sound detection under domain shifts,'' in \emph{Proc. EUSIPCO}, 2022.

\bibitem{lan2024hierarchical}
H.~Lan, Q.~Zhu, J.~Guan, Y.~Wei, and W.~Wang, ``Hierarchical metadata information constrained self-supervised learning for anomalous sound detection under domain shift,'' in \emph{Proc. ICASSP}, 2024.

\bibitem{guan2025disentangling}
J.~Guan, J.~Tian, Q.~Zhu, F.~Xiao, H.~Zhang, and X.~Liu, ``Disentangling hierarchical features for anomalous sound detection under domain shift,'' in \emph{Proc. ICASSP}, 2025.

\bibitem{harada2023first-shot}
N.~Harada, D.~Niizumi, Y.~Ohishi, D.~Takeuchi, and M.~Yasuda, ``First-shot anomaly sound detection for machine condition monitoring: {A} domain generalization baseline,'' in \emph{Proc. EUSIPCO}, 2023.

\bibitem{wilkinghoff2023design}
K.~Wilkinghoff, ``Design choices for learning embeddings from auxiliary tasks for domain generalization in anomalous sound detection,'' in \emph{Proc. ICASSP}, 2023.

\bibitem{wilkinghoff2025local}
K.~Wilkinghoff, H.~Yang, J.~Ebbers, F.~G. Germain, G.~Wichern, and J.~{Le Roux}, ``Local density-based anomaly score normalization for domain generalization,'' \emph{arXiv preprint arXiv:2509.10951}, 2025.

\bibitem{saengthong2024deep}
P.~Saengthong and T.~Shinozaki, ``Deep generic representations for domain-generalized anomalous sound detection,'' in \emph{Proc. ICASSP}, 2025.

\bibitem{ioffe2015batch}
S.~Ioffe and C.~Szegedy, ``Batch normalization: accelerating deep network training by reducing internal covariate shift,'' in \emph{Proc. ICML}, 2015.

\bibitem{carluzzi2017autodial}
F.~M. Carlucci, L.~Porzi, B.~Caputo, E.~Ricci, and S.~R. Bul{\`{o}}, ``{AutoDIAL: A}utomatic domain alignment layers,'' in \emph{Proc. ICCV}, 2017.

\bibitem{nichol2018first}
A.~Nichol, J.~Achiam, and J.~Schulman, ``On first-order meta-learning algorithms,'' \emph{arXiv preprint arXiv:1803.02999}, 2018.

\bibitem{snell2017prototypical}
J.~Snell, K.~Swersky, and R.~S. Zemel, ``Prototypical networks for few-shot learning,'' in \emph{Proc. NeurIPS}, 2017.

\bibitem{wang2023generalizing}
J.~Wang, C.~Lan, C.~Liu, Y.~Ouyang, T.~Qin, W.~Lu, Y.~Chen, W.~Zeng, and P.~S. Yu, ``Generalizing to unseen domains: {A} survey on domain generalization,'' \emph{{IEEE} Trans. Knowl. Data Eng.}, vol.~35, no.~8, 2023.

\bibitem{zhou2023domain}
K.~Zhou, Z.~Liu, Y.~Qiao, T.~Xiang, and C.~C. Loy, ``Domain generalization: {A} survey,'' \emph{{IEEE} Trans. Pattern Anal. Mach. Intell.}, vol.~45, no.~4, 2023.

\bibitem{dohi2022mimiidg}
K.~Dohi, T.~Nishida, H.~Purohit, R.~Tanabe, T.~Endo, M.~Yamamoto, Y.~Nikaido, and Y.~Kawaguchi, ``{MIMII} {DG:} {S}ound dataset for malfunctioning industrial machine investigation and inspection for domain generalization task,'' in \emph{Proc. DCASE}, 2022.

\bibitem{harada2023toyadmos2+}
N.~Harada, D.~Niizumi, D.~Takeuchi, Y.~Ohishi, and M.~Yasuda, ``{ToyADMOS2+}: New toyadmos data and benchmark results of the first-shot anomalous sound event detection baseline,'' in \emph{Proc. DCASE}, 2023.

\bibitem{niizumi2024toyadmos2sharp}
D.~Niizumi, N.~Harada, Y.~Ohishi, D.~Takeuchi, and M.~Yasuda, ``{ToyADMOS2\#: Y}et another dataset for the {DCASE2024} challenge task 2 first-shot anomalous sound detection,'' in \emph{Proc. DCASE}, 2024.

\bibitem{albertini2024imad-ds}
D.~Albertini, F.~Augusti, K.~Esmer, A.~Bernardini, and R.~Sannino, ``{IMAD-DS: A} dataset for industrial multi-sensor anomaly detection under domain shift conditions,'' in \emph{Proc. DCASE}, 2024.

\bibitem{chawla2002smote}
N.~V. Chawla, K.~W. Bowyer, L.~O. Hall, and W.~P. Kegelmeyer, ``{SMOTE:} {S}ynthetic minority over-sampling technique,'' \emph{J. Artif. Intell. Res.}, vol.~16, 2002.

\bibitem{ben-david2006analysis}
S.~Ben{-}David, J.~Blitzer, K.~Crammer, and F.~Pereira, ``Analysis of representations for domain adaptation,'' in \emph{Proc. NeurIPS}, 2006.

\bibitem{bardes2022vicreg}
A.~Bardes, J.~Ponce, and Y.~LeCun, ``{VICReg: V}ariance-invariance-covariance regularization for self-supervised learning,'' in \emph{Proc. ICLR}, 2022.

\bibitem{zhang2017mixup}
H.~Zhang, M.~Cisse, Y.~N. Dauphin, and D.~Lopez-Paz, ``Mixup: {B}eyond empirical risk minimization,'' in \emph{Proc. ICLR}, 2018.

\bibitem{sun2016deep}
B.~Sun and K.~Saenko, ``Deep {CORAL:} correlation alignment for deep domain adaptation,'' in \emph{Proc. ECCV}, 2016.

\bibitem{nam2022filteraugment}
H.~Nam, S.~Kim, and Y.~Park, ``Filteraugment: An acoustic environmental data augmentation method,'' in \emph{Proc. ICASSP}, 2022.

\bibitem{zhang2022towards}
H.~Zhang, Y.~Zhang, W.~Liu, A.~Weller, B.~Sch{\"{o}}lkopf, and E.~P. Xing, ``Towards principled disentanglement for domain generalization,'' in \emph{Proc. CVPR}, 2022.

\bibitem{ganin2016domain}
Y.~Ganin, E.~Ustinova, H.~Ajakan, P.~Germain, H.~Larochelle, F.~Laviolette, M.~Marchand, and V.~S. Lempitsky, ``Domain-adversarial training of neural networks,'' \emph{JMLR}, vol.~17, 2016.

\bibitem{lin2017focal}
T.~Lin, P.~Goyal, R.~B. Girshick, K.~He, and P.~Doll{\'{a}}r, ``Focal loss for dense object detection,'' in \emph{Proc. ICCV}, 2017.

\bibitem{wilkinghoff2025keeping}
K.~Wilkinghoff, H.~Yang, J.~Ebbers, F.~G. Germain, G.~Wichern, and J.~{Le Roux}, ``Keeping the balance: {A}nomaly score calculation for domain generalization,'' in \emph{Proc. ICASSP}, 2025.

\bibitem{wang2020continuously}
H.~Wang, H.~He, and D.~Katabi, ``Continuously indexed domain adaptation,'' in \emph{Proc. ICML}, 2020.

\bibitem{wang2022continual}
Q.~Wang, O.~Fink, L.~V. Gool, and D.~Dai, ``Continual test-time domain adaptation,'' in \emph{Proc. CVPR}, 2022.

\bibitem{mai2022explaining}
K.~T. Mai, T.~Davies, L.~D. Griffin, and E.~Benetos, ``Explaining the decision of anomalous sound detectors,'' in \emph{Proc. DCASE}, 2022.

\bibitem{tsubaki2023audio}
S.~Tsubaki, Y.~Kawaguchi, T.~Nishida, K.~Imoto, Y.~Okamoto, K.~Dohi, and T.~Endo, ``Audio-change captioning to explain machine-sound anomalies,'' in \emph{Proc. DCASE}, 2023.

\bibitem{gao2024learning}
J.~Gao, X.~Ma, and C.~Xu, ``Learning transferable conceptual prototypes for interpretable unsupervised domain adaptation,'' \emph{{IEEE} Trans. Image Process.}, vol.~33, 2024.

\bibitem{bobek2023towards}
S.~Bobek, S.~Nowaczyk, S.~Pashami, Z.~Taghiyarrenani, and G.~J. Nalepa, ``Towards explainable deep domain adaptation,'' in \emph{Proc. ECAI Workshops}, 2023.

\bibitem{zunino2021explainable}
A.~Zunino, S.~A. Bargal, R.~Volpi, M.~Sameki, J.~Zhang, S.~Sclaroff, V.~Murino, and K.~Saenko, ``Explainable deep classification models for domain generalization,'' in \emph{Proc. CVPR Workshops}, 2021.

\bibitem{aggarwal2017outlier}
C.~Aggarwal, \emph{Outlier Analysis}, 2nd~ed.\hskip 1em plus 0.5em minus 0.4em\relax Springer, 2017.

\bibitem{fujimura2025asdkit}
T.~Fujimura, K.~Wilkinghoff, K.~Imoto, and T.~Toda, ``{ASDKit}: A toolkit for comprehensive evaluation of anomalous sound detection methods,'' \emph{arXiv preprint arXiv:2507.10264}, 2025.

\end{thebibliography}

\end{document}